\documentclass[preprint,12pt,5p]{elsarticle}
\usepackage[italian,english]{babel}
\usepackage{hyperref}
\usepackage{ifpdf}
\usepackage{subfigure}

\usepackage{amssymb}
\usepackage{amsfonts}
\usepackage{epsf}
\usepackage{rotating}
\usepackage{graphicx}
\usepackage{amsmath}
\usepackage{fancyhdr}
\usepackage{lineno}

\usepackage{babel}
\usepackage{graphics}
\usepackage{pstricks}
\usepackage{color}
\usepackage{multirow}

\def\beq{\begin{equation}}
\def\eeq{\end{equation}}
\def\bea{\begin{eqnarray}}
\def\eea{\end{eqnarray}}
\def\nn{\nonumber}
\def\bei{\begin{itemize}}
\def\eei{\end{itemize}}
\def\bmat{\begin{matrix}}
\def\emat{\end{matrix}}
\def\={\,=\,}
\def\+{\,+\,}
\def\-{\,-\,}

\def\nn{\nonumber}
\def\gev{\ensuremath{\mathrm{Ge\kern -0.1em V}}}
\begin{document}

%\begin{frontmatter}

%%%%%%%%%%%%%%%%%%%%%%%%%%%%%%%%%%%%%%%%%%%%%
\title{Scrutinizing Right-Handed Neutrino portal Dark Matter \\ with Yukawa effect}

\author[a]{Priyotosh Bandyopadhyay}
\ead{bpriyo@iith.ac.in} 

\author[b]{Eung Jin Chun}
\ead{ejchun@kias.re.kr} 

\author[c]{Rusa Mandal}
\ead{Rusa.Mandal@ific.uv.es}

\author[d]{Farinaldo S. Queiroz}
\ead{farinaldo.queiroz@iip.ufrn.br}

\address[a]{Indian Institute of Technology Hyderabad, Kandi,  Sangareddy-502287, Telengana, India}
\address[b]{Korea Institute for Advanced Study, Seoul 130-722, Republic of Korea}

\address[c]{The 
	Institute  of Mathematical Sciences, HBNI, Taramani, Chennai 600113, India \\ and \\ IFIC, Universitat de Val$\grave{e}$ncia-CSIC, Apt. Correus 22085, E-46071 Val$\grave{e}$ncia, Spain}

\address[d]{International Institute of Physics, Universidade Federal do Rio Grande do Norte, Campus Universit\'ario, Lagoa Nova, Natal-RN 59078-970, Brazil}

%\preprint{ IITH-PH-0002/17}
	
%	\hspace*{11.27cm} IMSc/2017/07/05}

%%%
\begin{abstract}
Analyzing the neutrino Yukawa effect in the freeze-out process of a generic dark matter candidate with right-handed neutrino portal, we identify the parameter regions satisfying the observed dark matter relic density as well as the current Fermi-LAT and H.E.S.S. limits and the future CTA reach on gamma-ray signals. In this scenario the dark matter couples to the Higgs boson at one-loop level and thus could be detected by spin-independent nucleonic scattering for a reasonable range of the relevant parameters. 
	
\end{abstract}

\begin{keyword}
	Right-Handed Neutrino \sep Dark Matter \sep Direct and Indirect Detection
	
	
	
\end{keyword}

%\end{frontmatter}

\maketitle
\flushbottom
%%%%%%%%%%%%%%%%%%%%%%%%%%

%%%%%%%%%%%%%%%%%%%%%%%%%%%
\section{Introduction}
%%%%%%%%%%%%%%%%%%%%%%%%%%%

The smallness of neutrino masses may be explained by the presence of right-handed neutrinos (RHNs) with
large Majorana mass realizing the seesaw mechanism \cite{rhn}. 
It is conceivable that a dark matter (DM) candidate couples to a RHN and thus 
its pair-annihilation to a RHN pair is responsible for the DM freeze-out. 
Such a situation can be realized specifically when RHNs are introduced 
in association with an extended ($B-L$) gauge symmetry \cite{Bandyopadhyay:2011qm,Bandyopadhyay:2017bgh}. In this scenario, an interesting feature arises in the process of DM thermal freeze-out. 
Due to a tiny neutrino Yukawa coupling  of a RHN with lepton and Higgs doublet, the RHN may not be fully thermalized and thus the observed DM relic density can be achieved by the DM annihilation rate different from the standard freeze-out value \cite{Bandyopadhyay:2011qm,Bandyopadhyay:2017bgh}. 
Such a feature has been realized  also in various scenarios \cite{Dror:2016rxc,Okawa:2016wrr,Kopp:2016yji}.

The RHN as a portal to DM was suggested in a simple setup assuming  the coupling
$N \chi \phi$ where a fermion $\chi$ or a scalar $\phi$ can be a DM candidate \cite{posp07}, 
and studied extensively in Refs.~\cite{falk09,gonz16,esc16,tang16,camp17,batell17,Chianese:2018dsz}. 
In this paper, we explore the enlarged parameter space including  
the RHN Yukawa  effect to investigate how it is constrained by 
the thermal DM relic density, direct and indirect detections. 
We will assume that DM is the fermion field $\chi$, and thus the nucleon-DM scattering arises at one-loop through the $\phi$-$\phi$-Higgs coupling.

The rest of the paper is organized as follows. In Sec.~\ref{sec:relic}, after describing the RHN portal structure with a fermionic DM candidate, we discuss the impact of neutrino Yukawa couplings to the thermal freeze-out condition of the DM pair annihilation to a RHN pair. 
This allows us to identify parameter regions satisfying the observed DM relic density, 
which are constrained by indirect detection experiments. Applying the latest Fermi-LAT and H.E.S.S. data on gamma-ray signals, produced by RHN decays in our scenario, we put combined constraints on the model parameter space in Sec.~\ref{sec:indirect}.
In Sec.~\ref{sec:Direct}, we consider a direct detection process arising from one-loop induced DM-DM-Higgs coupling and limits from the recent data and future sensitivity on spin-independent (SI) DM scatterings.
Finally we conclude in Sec.~\ref{sec:conclusion}.

%%%%%%%%%%%%%%%%%%%%%%%%%%%%
\section{DM freeze-out including neutrino Yukawa effect} \label{sec:relic}
%%%%%%%%%%%%%%%%%%%%%%%%%%%%

Let us consider the simplistic scenario for a RHN-portal DM based on the Type-I seesaw. The Lagrangian of such construct will contain the following new terms
\begin{align}
\label{eq:Lag}
-\mathcal{L} \subset& \frac{1}{2} m_0^2 \phi^2 +\kappa \phi^2 |H|^2 + \Big\{ \frac{1}{2} m_\chi \chi \chi + \frac{1}{2} m_N NN \nn \\
 +& y_N LHN + \lambda N \chi \phi + {\rm h.c.} \Big\}.
\end{align}
Here $L$ and $H$ are the SM $SU(2)_L$ doublets and $N$ is a Majorana fermion (RHN). 
There are two new fields in the dark sector: a real scalar $\phi$ and  a Majorana fermion $\chi$ which are singlets under the SM gauge group. For the stability of a DM candidate, we assigned, e.g., a $Z_2$ parity under which the dark sector fields are odd.
After the electroweak symmetry breaking, $H=(v+h)/\sqrt{2}$, we get the scalar mass $m_\phi^2=m_0^2 + \kappa v^2$ and the $h$-$\phi$-$\phi$ coupling $\kappa v$.

There are two couplings $\lambda$ and  $\kappa$  
which connect the dark sector ($\phi$ and $\chi$) to the visible sector.
When $\phi$ is a thermal DM candidate, the Higgs portal coupling $\kappa$ plays an important role.
In this case,  the parameter space is highly constrained by various considerations including the latest XENON1T result  \cite{Athron:2018ipf}. The RHN-portal process, $\phi\phi \to NN$ through the $t$-channel exchange of $\chi$, can also be operative to produce the right thermal relic density. Notice that a similar situation was studied in Ref.~\cite{Bandyopadhyay:2011qm} where $\phi$ corresponds to a right-handed sneutrino DM. 
In this paper, we concentrate on the fermion $\chi$ as a DM candidate. 
Our results on the RHN-portal property can also be applied to the case of the scalar $\phi$ as a DM candidate.

\begin{figure}[h]
	\begin{center}
		\includegraphics[width=1.\linewidth]{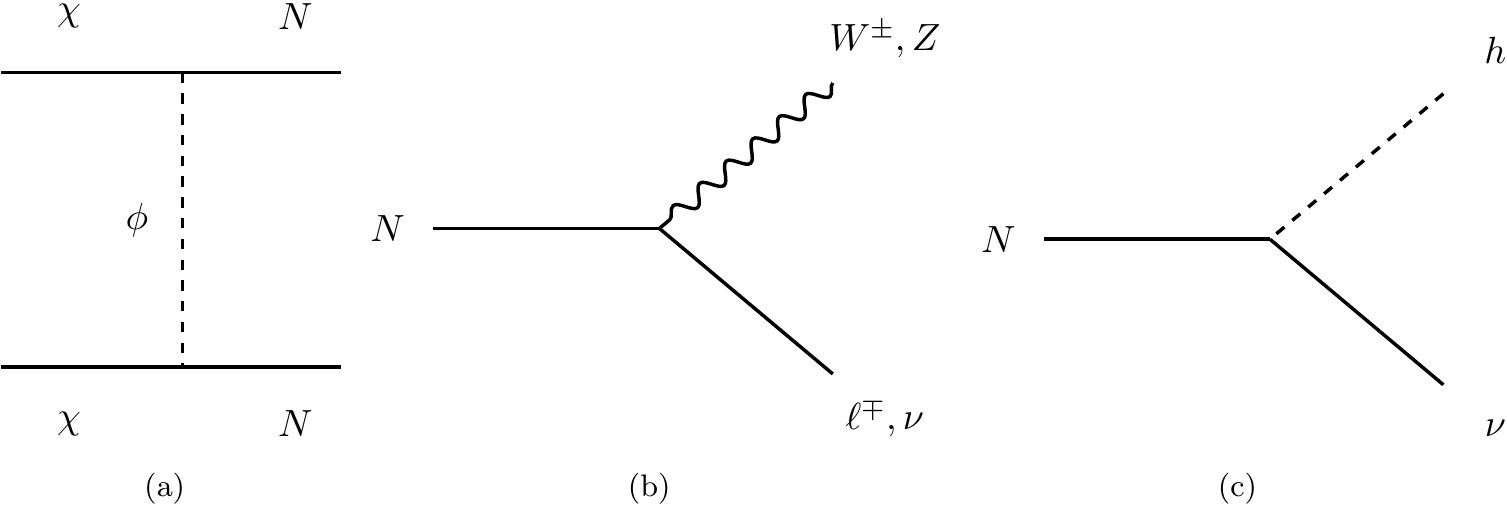}
		\caption{The Feynman diagrams for the DM particle $\chi$ annihilation to RHN $N$ pair (a) and the decay of $N$ to SM particles (b), (c) are shown. }\label{dia:relic}
	\end{center}
\end{figure}

When $\chi$ is lighter than $\phi$, it becomes a viable DM candidate. 
For $m_N<m_\chi$, the DM particle $\chi$ can annihilate to the RHN pair via a $t$-channel exchange of $\phi$ (Fig.~\ref{dia:relic}(a)). The thermal average annihilation cross section is given by,
\begin{align}\label{sigv}
\hspace*{-0.2cm}\langle\sigma v \rangle_{\chi\chi\to NN} = \frac{\lambda^4 \left(m_\chi + m_N \right)^2}{16 \pi \left( m_\chi^2 +m_\phi^2 -m_N^2 \right)^2} \left(1- \frac{m_N^2}{m_\chi^2}\right)^{1/2}\!\!\!\!\!.
\end{align}
There are other relevant annihilation processes like $\phi \phi \to \chi \chi\,(N,N)$, $\phi \phi \to $ SM particles and co-annihilation channel $\chi\phi\to N \to$ SM particles which can contribute in the evaluation of DM number density. We quote these expressions in ~\ref{sec:appendix}. We notice that the co-annihilation channel is suppressed by two reasons; firstly the tiny Yukawa coupling $y_N$, secondly the choice of parameter space away from the resonant $N$ production, and thus has insignificant effect in the freeze out mechanism.
We start with the coupled Boltzmann equations written in terms of the variable $Y_i\equiv n_i/s$, describing the actual number of particle $i$ per comoving volume, where $n_i$ being the number density, $s$ is the entropy density of the Universe, and the variable $x\equiv m_\chi/T$. 
The Boltzmann equations relevant for our study are
\begin{align}
\label{eq:dYDM}
\frac{d Y_\chi}{dx}=& -\frac{1}{x^2} \frac{s(m_\chi)}{H(m_\chi)} \langle \sigma v \rangle_{\chi\chi\to NN}\!\! \left(\!\!Y_\chi^2 \!-\! \left(\frac{Y_\chi^{\text{eq}}}{Y_N^{\text{eq}}}\right)^{\!\!\!2}Y_N^2\!\right) \nn \\ 
+ \frac{1}{x^2}& \frac{s(m_\chi)}{H(m_\chi)} \langle \sigma v \rangle_{\phi\phi\to\chi\chi}\! \left(\!\!Y_\phi^2 \!-\! \left(\!\frac{Y_\phi^{\text{eq}}}{Y_\chi^{\text{eq}}}\right)^{\!\!\!2}\!Y_\chi^2\!\right)\!, \\
\label{eq:dYphi}
\frac{d Y_\phi}{dx}=& - \frac{1}{x^2} \frac{s(m_\chi)}{H(m_\chi)} \langle \sigma v \rangle_{\phi\phi\to\chi\chi} \left(\!Y_\phi^2 \!-\! \left(\!\frac{Y_\phi^{\text{eq}}}{Y_\chi^{\text{eq}}}\right)^{\!\!\!2}Y_\chi^2\right) \nn \\
-&\frac{1}{x^2} \frac{s(m_\chi)}{H(m_\chi)} \langle \sigma v \rangle_{\phi\phi\to NN} \left(\!Y_\phi^2 \!-\! \left(\frac{Y_\phi^{\text{eq}}}{Y_N^{\text{eq}}}\right)^{\!\!\!2} Y_N^2\!\right) \nn \\
-&\frac{1}{x^2} \frac{s(m_\chi)}{H(m_\chi)} \langle \sigma v \rangle_{\phi\phi\to {\rm SM}} \left(Y_\phi^2 - {Y_\phi^{\text{eq}}}^2\right), \\
\label{eq:dYN}
\frac{d Y_N}{dx} =& \frac{1}{x^2} \frac{s(m_\chi)}{H(m_\chi)} \langle \sigma v \rangle_{\chi\chi \to NN} \left(\!Y_\chi^2 \! -\! \left(\frac{Y_\chi^{\text{eq}}}{Y_N^{\text{eq}}}\right)^{\!\!\!2} Y_N^2\!\right) \nn \\
+& \frac{1}{x^2} \frac{s(m_\chi)}{H(m_\chi)} \langle \sigma v \rangle_{\phi\phi\to NN} \left(\!Y_\phi^2 \!-\! \left(\frac{Y_\phi^{\text{eq}}}{Y_N^{\text{eq}}}\right)^{\!\!\!2}Y_N^2\!\right) \nn \\
-&\frac{\Gamma}{H(m_\chi)} x \left(Y_N -Y_N^{\text{eq}}\right).
\end{align}

\begin{figure*}[!h]
	\begin{center}
		\mbox{\hskip -20 pt \subfigure[]{\includegraphics[width=0.45\linewidth]{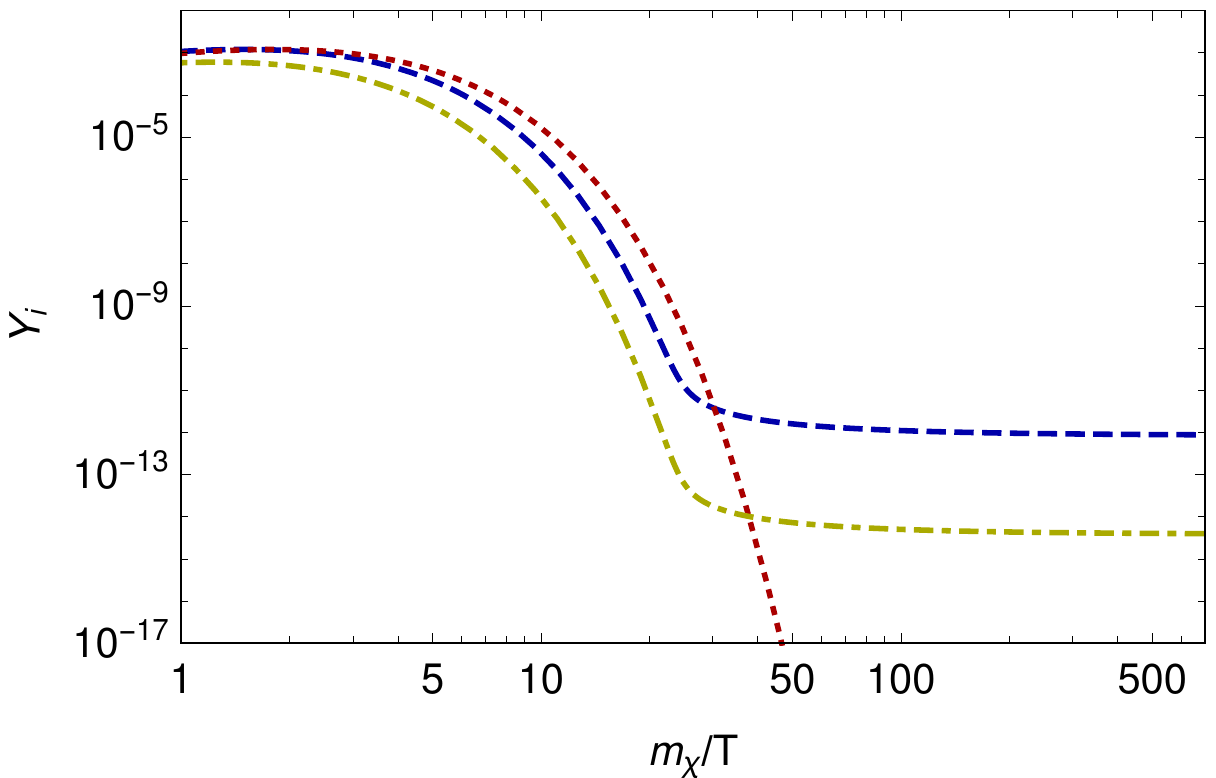}}
			\subfigure[]{\includegraphics[width=0.45\linewidth]{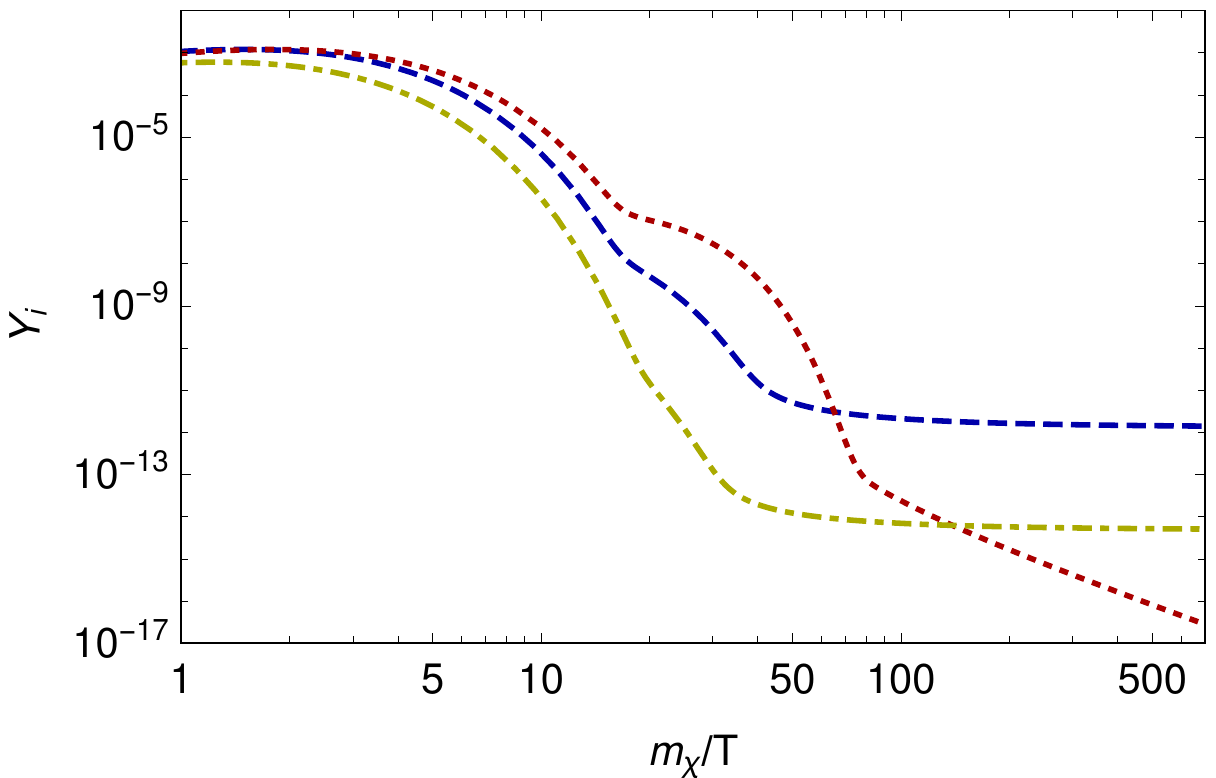}}}
		\mbox{\hskip -20 pt
			\subfigure[]{\includegraphics[width=0.45\linewidth]{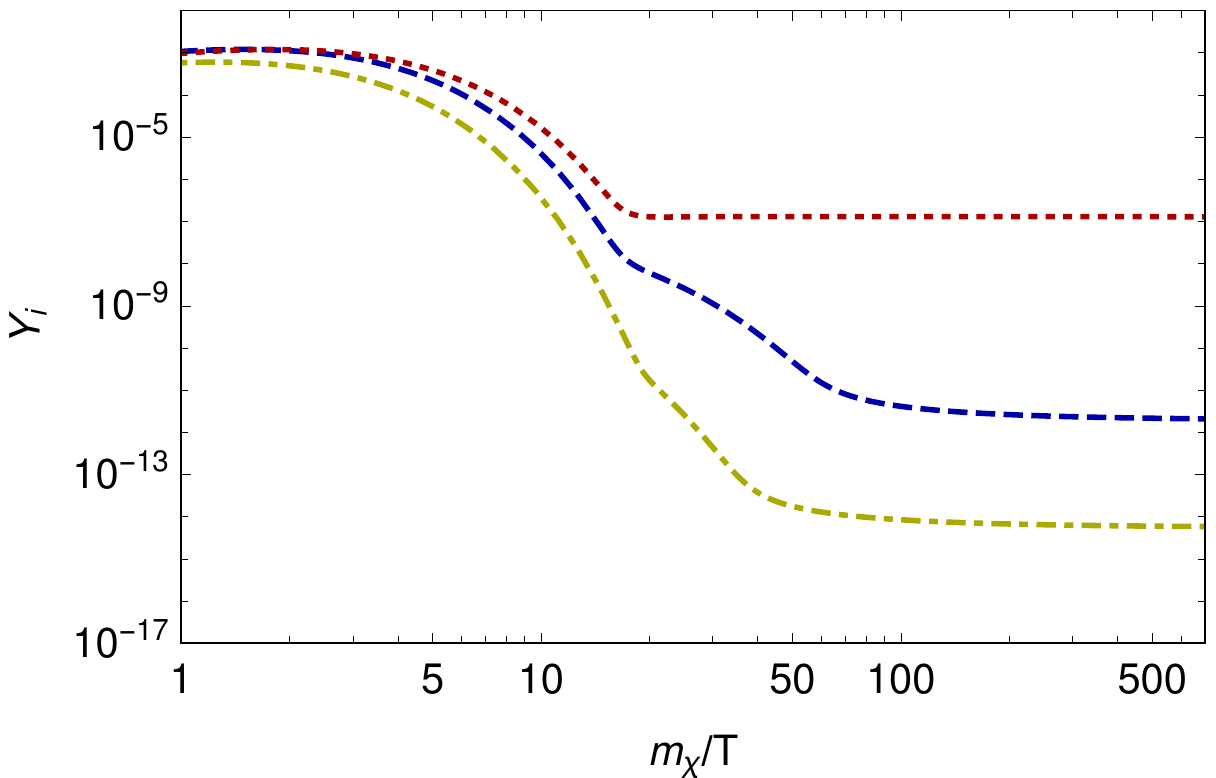}}
			\subfigure[]{\includegraphics[width=0.45\linewidth]{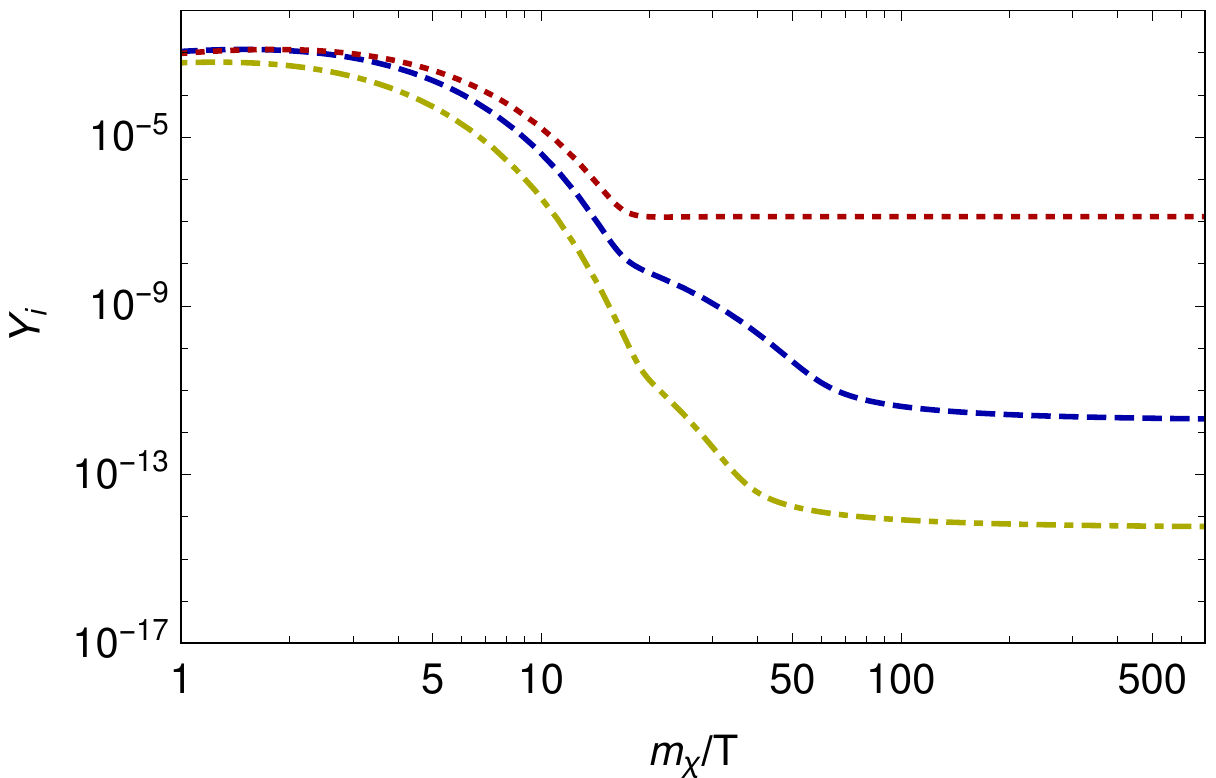}}}
		\caption{The actual number of $\chi,~\phi$ and $N$ per comoving volume are shown in blue dashed, green dot-dashed and red dotted curves, respectively. The panels from (a) to (d) are obtained by solving the coupled Boltzmann equations (Eqs.~\eqref{eq:dYDM} -- \eqref{eq:dYN}) with the total decay width $\Gamma$ of $N$ as $10^{-10}\, \gev$, $10^{-15}\,\gev$, $10^{-20}\,\gev$ and $0\, \gev$, respectively. The effect of decay term is prominent from the plots. The masses of  $\chi$, $N$, $\phi$ are assumed to follow $m_\chi=n\, m_N= 1/n\, m_\phi$ with $n=1.2$, $m_N=300$\,GeV and the couplings $\lambda=0.4$, $\kappa=1$. The observed relic density is satisfied in panel (b) with $\Gamma= 10^{-15}$\,GeV.}
		\label{fig:density}
	\end{center}
\end{figure*}

\noindent
The entropy density $s$ and Hubble parameter $H$ at the DM mass is
$$ 
s(m_\chi)= \frac{2 \pi^2 }{45} g_*\, m_\chi^3, \quad H(m_\chi)= \frac{\pi}{\sqrt{90}} \frac{\sqrt{g_*}}{M^r_{pl}} m_\chi^2, $$  where $M^r_{pl}= 2.44\times {10}^{18}\gev$ is the reduced Planck mass and $Y_N^{\text{eq}}$ is the equilibrium number density of $i$-th particle given by
\begin{align}
\label{eq:Yi_eq}
Y_i^{\text{eq}}&\equiv\frac{n_i^{\text{eq}}}{s} =\frac{45}{2\pi^4} \sqrt{\frac{\pi}{8}}\left( \frac{g_i}{g_*}\right) \left({\frac{m_i}{T}}\right)^{3/2} e^{-\frac{m_i}{T}} \nn \\
&\simeq 0.145 \left( \frac{g_i}{100}\right)  \left( \frac{m_i}{m_\chi}\right)^{3/2} x^{3/2} e^{-\frac{m_i}{m_\chi} x}.
\end{align}
Here in the last line of Eq.~\eqref{eq:Yi_eq} we use the effective number of relativistic degrees of freedom $g_*\simeq100$ and the internal degrees of freedom $g_{\chi,N}=2$ for the two Majorana particles $\chi$, $N$ and $g_\phi=1$ for $\phi$ being the real scalar.
The first terms on the right-hand side of Eqs.~\eqref{eq:dYDM} and \eqref{eq:dYN} denote the forward and backward reactions of $\chi\chi$ to $NN$ through $t$-channel $\phi$ exchange shown in Fig.~\ref{dia:relic}(a). It can be seen from Eq.~\eqref{eq:Lag} that the Yukawa interaction of the right-handed neutrino allows it to decay to SM particles via the mixing with the SM neutrinos proportional to the coupling $y_N$. The third term of Eq.~\eqref{eq:dYN} describes the decay and the inverse decay of $N$ shown in Fig.~\ref{dia:relic}(b) and (c) where $\Gamma$ being the total decay width of $N$. Below we quote the expressions of the partial decay widths of $N$ to three possible channels $h\nu$, $\ell^\pm W^\mp$ and $Z\nu$, respectively.
\begin{align}
\label{eq:Ndecayh}
\Gamma(N\to h \nu)=&\Gamma(N\to h \bar{\nu}) \nn \\
	=& \frac{y_N^2 m_N}{64\pi} \left(1- \frac{m_h^2}{m_N^2}\right)^2, \\ 
\Gamma(N\to \ell^- W^+)= &\Gamma(N\to \ell^+ W^-) \nn \\
	= \frac{y_N^2 m_N}{32\pi} & \left(1- \!\frac{m_W^2}{m_N^2}\right)^2\!\! \left(1+ 2 \frac{m_W^2}{m_N^2}\right)\!, \\ 
\label{eq:NdecayZ}
\Gamma(N\to Z \nu ) = &\Gamma(N\to Z \bar{\nu}) \nn \\
	=  \frac{y_N^2 m_N}{64\pi} & \left(1- \frac{m_Z^2}{m_N^2}\right)^2 \!\! \left(1+ 2 \frac{m_Z^2}{m_N^2}\right).
\end{align}
The relic abundance of the DM candidate $\chi$ can be evaluated by,
\begin{align}
\label{eq:relic}
\Omega h^2 = \frac{m_\chi s_0 Y_\chi(\infty)}{\rho_c/h^2},
\end{align}
where $s_0=2890$ cm$^{-3}$ is the current entropy density of the Universe and $\rho_c/h^2=1.05\times 10^{-5}\, \gev/ $cm$^3$ is the critical density. $Y_\chi(\infty)$ is the asymptotic value of the actual number of $\chi$ per comoving volume obtained from numerical solutions of the above Boltzmann equations. We illustrate the effect of decay and inverse decay of RHN in the evaluation of DM density, for a benchmark case, in Fig.~\ref{fig:density}. It can be seen that, in this case, the contribution of scalar DM $\phi$ to relic density is negligible compared to the Majorana fermion $\chi$.

Depending on the flavor structure of the Yukawa coupling $y_N$, the RHN decays differently to each lepton flavor,
which will lead to a different prediction for indirect detection.
For our analysis of indirect detection, we will assume $N$ decaying equally to three lepton flavors. 

%%%%%%%%%%%%%%%%%%%%%%%%%%%%%%%%%%
\section{Indirect Detection}  \label{sec:indirect}
%%%%%%%%%%%%%%%%%%%%%%%%%%%%%%%%%%

\begin{figure*}[!h]
	\begin{center}
		\includegraphics[width=0.49\linewidth]{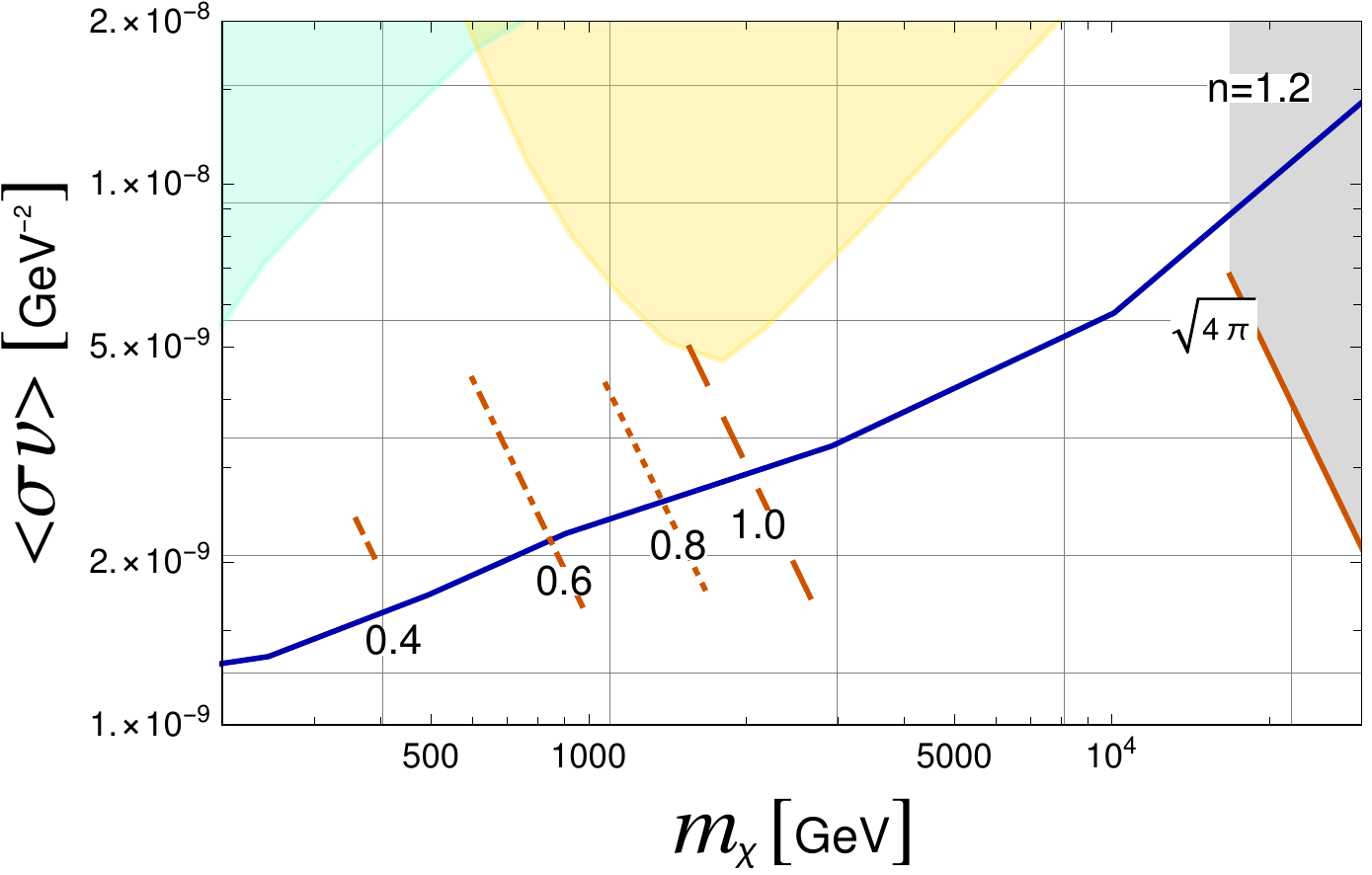}
		\includegraphics[width=0.49\linewidth]{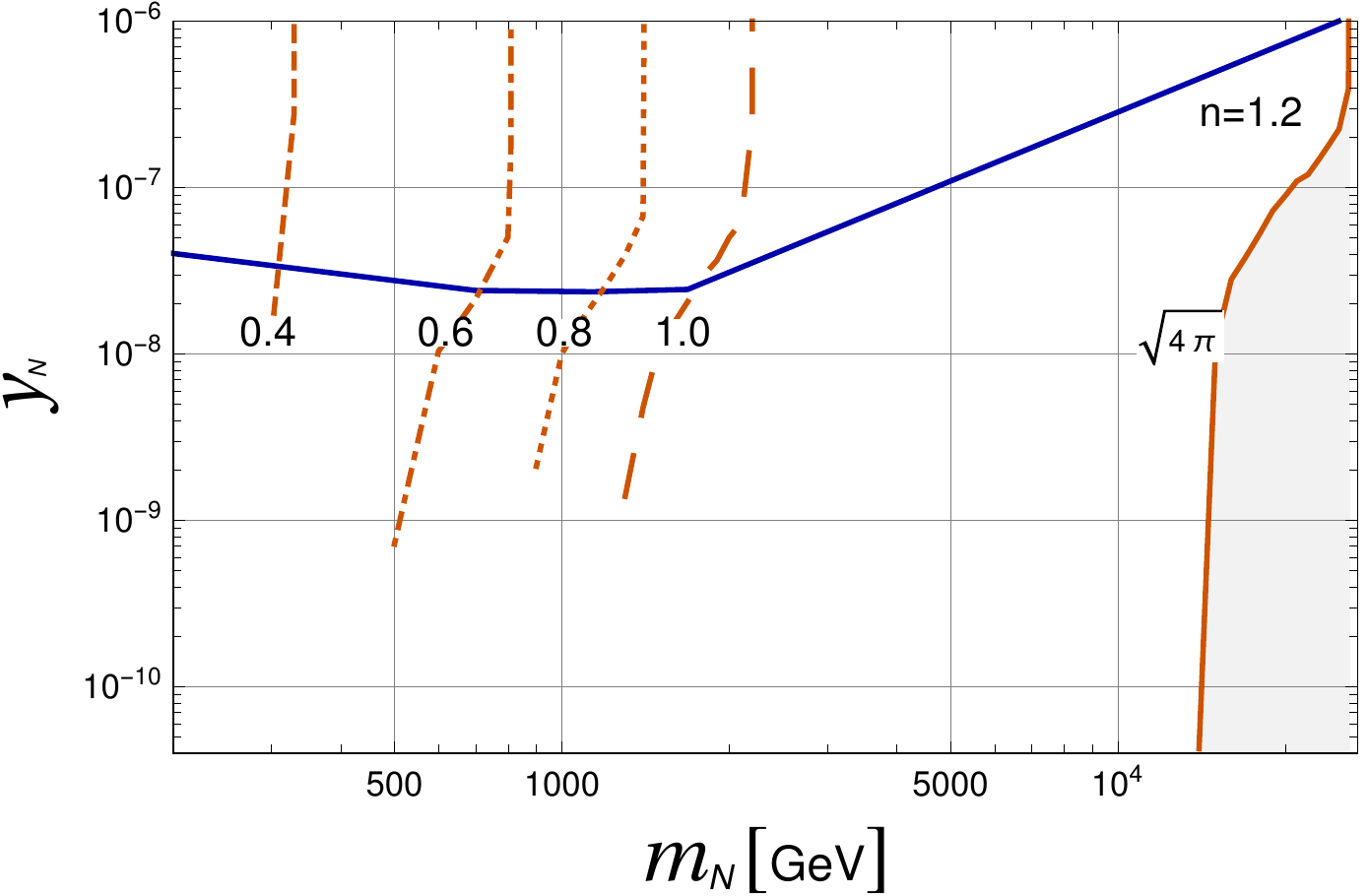}
		\caption{The left and right panels show the allowed parameter space in the plane of  
			($m_\chi$ $\langle \sigma v\rangle$) and  ($m_N$,$y_N$), respectively. 
			The observed relic density is obtained for the DM coupling $\lambda$=0.4 (dashed), 0.6 (dot-dashed), 0.8 (dotted), 1.0 (long-dashed) and $\sqrt{4\pi}$ (solid).
			The green and yellow shaded regions are excluded by Fermi-LAT(at 90\% C.L.) and H.E.S.S.(at 95\% C.L.) data, respectively. The blue solid curve represents future bound from CTA where the region above(below) will be excluded at 90\% C.L. for left(right) panel. 
			The gray region is forbidden by perturbativity limit. 
			The masses of  $\chi$, $N$, $\phi$ are assumed to follow $m_\chi=n\, m_N= 1/n\, m_\phi$ with $n=1.2$, $\kappa=1$,
			and the RHN is assumed to decay equally to each lepton flavor.
		}\label{fig:sigV}
	\end{center}
\end{figure*}

Here we would like to mention that the RHN-portal models can be probed by indirect detection experiments. The annihilation of DM pair to RHNs, which then decay through weak interactions induced by  active-sterile neutrino mixing, leads to gamma-ray signals that can be probed by experiments such as Fermi-LAT and H.E.S.S. telescopes \cite{camp17,batell17}. In our work we employed the receipt described in \cite{camp17} to find the H.E.S.S. bounds and the results from \cite{Folgado:2018qlv} for the Fermi-LAT bound on the dark matter annihilation cross section for the $\chi\chi \to N N$ process which is depicted in Fig.~\ref{fig:sigV}. We emphasize that H.E.S.S. and CTA limits rely on the current (projected) sensitivity to gamma-ray emission stemming from the Galactic Center. Since no excess has been observed, stringent constraints have been placed on the dark matter annihilation cross section. It is clear from the figure that the CTA limit is more constraining and this is a direct result of the CTA array containing Large, Medium and Small-Sized Telescopes that will significantly strengthen CTA sensitivity to dark matter models 
\cite{Acharya:2017ttl}. We focus our discussion on the benchmark scenario where $m_\chi=n\, m_N= 1/n\, m_\phi$. 

The left panel of Fig.~\ref{fig:sigV} in the  $\langle \sigma v\rangle-m_\chi$ plane shows the lines satisfying observed relic abundance by Planck data $\Omega h^2 = 0.1199 \pm 0.0027$ ~\cite{Ade:2015xua} achieved for different values of the coupling $\lambda$. The green and yellow shaded regions depict 90\% C.L. limit on annihilation cross section from Fermi-LAT~\cite{Folgado:2018qlv} and 95\% C.L. bound from H.E.S.S. data~\cite{camp17}, respectively. The right-panel shows the corresponding situation in the $m_N-y_N$ plane. One can observe an important feature that given a fixed value of  $\lambda$, the observed relic can be obtained for quite extended ranges of the DM mass  $m_\chi$ by changing the neutrino Yukawa coupling $y_N$, {\em viz} controlling the decay width $\Gamma$. This parameter space is currently allowed by the limits from indirect detection experiments however can be probed by the projected bound from CTA in future. The system of the coupled Boltzmann equations, \eqref{eq:dYDM} and \eqref{eq:dYN}, reduces to the conventional one where the RHN is assumed to be in thermal equilibrium. This is realized when $\langle \sigma v \rangle \simeq 2 \times 10^{-9}\, \gev^2$ and  the result becomes independent of $y_N$, which is nicely depicted  in the right panel. 
The gray shaded region is forbidden by the perturbativity limit on $\lambda$. 
For higher values of $n$, the parallel lines for $y_N \geq 10^{-7}$ in the left panel of Fig.~\ref{fig:sigV} would be satisfied for higher values of $\lambda$ for a given $m_N$. This is due to the fact that an increase in $n$ decreases $\langle\sigma v\rangle$, which can be read from Eq.~\ref{sigv}.

%%%%%%%%%%%%%%%%%%%%%%%%%%%%%%%%%%%%%%%%%%
\section{Direct detection} \label{sec:Direct}
%%%%%%%%%%%%%%%%%%%%%%%%%%%%%%%%%%%%%%%%%%%%
\begin{figure}[h]
	\begin{center}
		\includegraphics[width=0.5\linewidth]{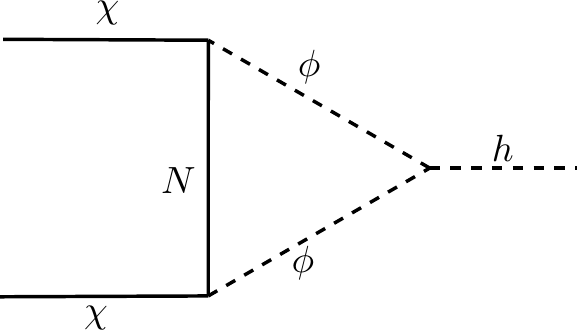}
		\caption{The interaction of the DM $\chi$ with the Higgs $h$ induced at one-loop level. }\label{dia:hDD}
	\end{center}
\end{figure}
\begin{figure}[t]
	\begin{center}
		\includegraphics[width=1\linewidth]{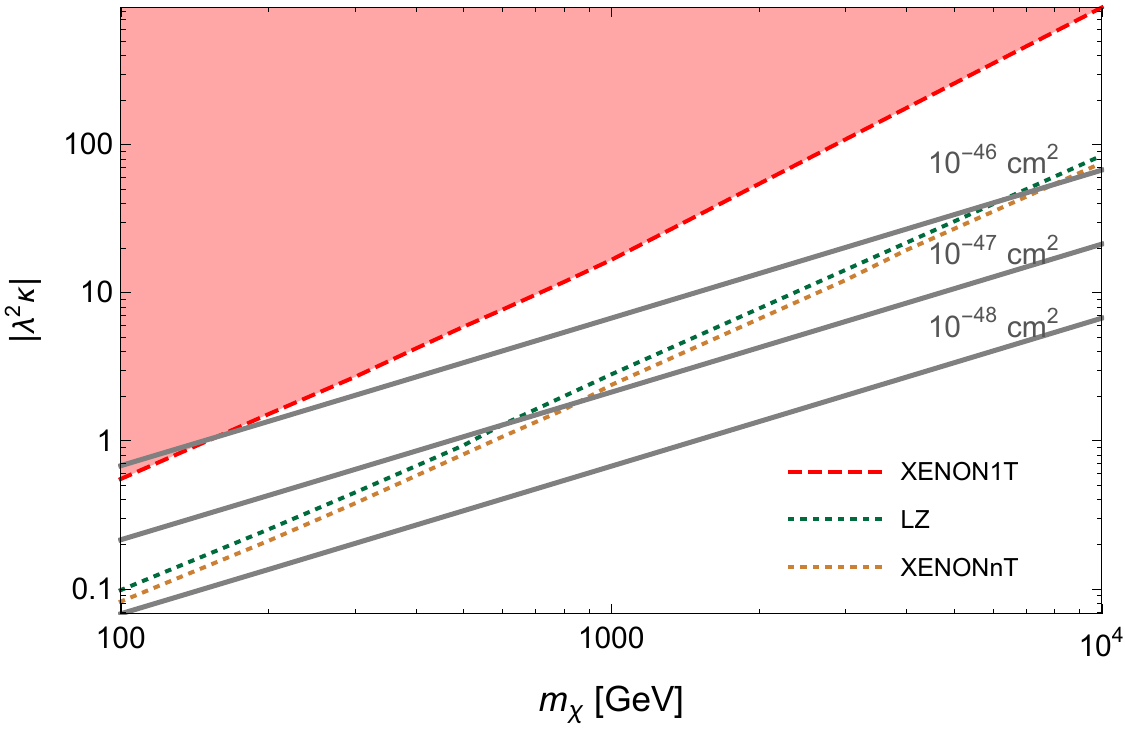}
		\caption{The contour plot for direct detection cross-section through a loop induced $h$-$\chi$-$\chi$ coupling is shown in $m_\chi- |\lambda^2\kappa|$ plane. 
			The 2018 XENON1T bound~\cite{Aprile:2018dbl} is shown by the red-dashed curve.
			The green- and orange-dotted curves are the expected bounds from LZ \cite{Mount:2017qzi}
			and XENONnT~\cite{Aprile:2015uzo} experiments, respectively. The region above the mentioned curves are excluded at 90\% confidence level.
		}\label{fig:DirectD}
	\end{center}
\end{figure}

Notice that the model contains no tree-level coupling of the  fermionic  DM to the Higgs boson, but an effective $h$-$\chi$-$\chi$ coupling arises from the one-loop diagram shown in  Fig.~\ref{dia:hDD}:
\begin{align} \label{eq:hDD}
-{\cal L}_{h\chi\chi}=& \kappa' h \bar\chi \chi ~~\mbox{where} \nn \\  
\kappa' \equiv& {\lambda^2 \kappa v \over 16\pi^2}\, {m_\chi c_1(x) -m_N c_0(x) \over m_\phi^2},
\end{align}
and $c_{1,0}(x)$ are loop-functions of $x\equiv m_N^2/m_\phi^2$ given by
\begin{eqnarray}
c_1(x) &=& \frac{1-4x+3x^2-2x^2 \ln x}{2(1-x)^3}, \nn \\
c_0(x) &=& \frac{1-x+x\ln x}{(1-x)^2} . \nonumber
\end{eqnarray}
The induced $h$-$\chi$-$\chi$ coupling $\kappa'$ (Eq.~\eqref{eq:hDD})  controls 
the SI nucleonic cross-section
\begin{equation}
\sigma_{\rm SI} = \frac{4}{\pi} \mu_r^2
\left({ \kappa' g_{nnh} \over m_h^2 }\right)^2,
\end{equation}
where $\mu_r=m_\chi m_n/(m_\chi+m_n)$ is the reduced mass  and $g_{nnh} \approx 0.0011$ is the nucleon-Higgs coupling.  The measurements of DM-nucleon SI cross section constrain the effective Higgs-DM coupling stringently and the result is depicted in Fig.~\ref{fig:DirectD} which shows the latest bound from XENON1T 2018 result \cite{Aprile:2018dbl} and the future limits from LZ \cite{Mount:2017qzi} and XENONnT~\cite{Aprile:2015uzo} experiments. The region above the mentioned curves are excluded at 90\% confidence level. 

It can be seen that the latest data from XENON1T experiment excludes $|\lambda^2\kappa|\ge \mathcal{O}(1)$ for $m_\chi\le 150\,$GeV and the future sensitivity of XENONnT can rule out such value of $|\lambda^2\kappa|$ up to $600\,{\rm GeV}$ DM mass.
As the direct detection process arises at one-loop level with an additional coupling $\kappa$ irrelevant for the DM annihilation, there remains a wide range of parameter space to be probed by both direct and indirect detections.

%%%%%%%%%%%%%%%%%%%%%%%%%%%%%%%%%%%%%
\section{Conclusion} \label{sec:conclusion}
%%%%%%%%%%%%%%%%%%%%%%%%%%%%%%%%%%%%%%

The dark sector may possibly be connected to the visible sector through heavy Majorana RHNs
which are introduced to explain the observed neutrino masses and mixing.  
Assuming a fermionic DM candidate which pair-annihilates to a RHN pair, 
we performed a comprehensive analysis of the parameter space considering the neutrino Yukawa effect in 
the thermal freeze-out process and imposing the current results of indirect and direct detection experiments.
When the neutrino Yukawa coupling is too small to maintain the RHN in full thermal equilibrium, the DM annihilation cross-section needs to be larger than the standard freeze-out value to obtain the observed relic density. However, the allowed parameter region is
quite limited and well below the current limits from Fermi-LAT and H.E.S.S. telescopes detecting gamma ray signals.  
The CTA will be able to probe a large part of the region as shown in Fig.~\ref{fig:sigV}.
In this scenario, a DM-Higgs coupling arises at one loop and thus could be probed by direct detection experiments through spin-independent scattering. The 2018 XENON1T bound and future limits are illustrated in Fig.~\ref{fig:DirectD}.

\section*{Acknowledgments}

We thank Christoph Weniger for the discussion and encouragement.
EJC is supported by the NRF grant funded by the Korea government (MSIP) (No. 2009-
0083526) through KNRC at Seoul National University.
FSQ thanks the financial support from UFRN, MEC and  ICTP-SAIFR FAPESP grant 2016/01343-7. 
The work of RM has been supported in part by Grants No. FPA2014-53631-C2-1-P, FPA2017-84445-P and SEV-2014-0398 (AEI/ERDF, EU) and by PROMETEO/\\2017/053 (GV, ES).
\appendix

\section{}
\label{sec:appendix}
In this section we provide the expressions for cross sections involved in the coupled Boltzmann equations (Eqs.~\eqref{eq:dYDM} -- \eqref{eq:dYN}). The scalar DM particle $\phi$ can annihilate to $\chi(N)$ pair via a $t$-channel exchange of $N(\chi)$ and the thermally averaged cross section is given in Eq.~\eqref{eq:sigvXX}(\eqref{eq:sigvNN}). The $\phi$ pair can also annihilate to the SM particles where the dominant channels are $\phi\phi\to hh$ and $\phi\phi\to h \to t\bar{t},~WW,~ZZ$ where $h$ is the SM Higgs boson. The process $\phi\phi\to hh$ combines three contributions as shown in Eq.~\eqref{eq:PPhh}; contact 4-point interaction (first term), $s$-channel Higgs exchange (second term) and $t$-channel $\phi$ exchange (third term). All three channels written in Eq.~\eqref{eq:PPSM} proceed through a $s$-channel Higgs exchange and hence are less dominant far from the resonant Higgs production. We use these expressions in solving the Boltzmann equations (Eqs.~\eqref{eq:dYDM} -- \eqref{eq:dYN}) in Secs. \ref{sec:relic} and \ref{sec:indirect}.
\begin{align}
\label{eq:sigvXX}
&\langle\sigma v\rangle_{\phi\phi\to\chi\chi} =\frac{\lambda^4 \left(m_\chi + m_N \right)^2}{2 \pi \left( m_\phi^2-m_\chi^2 + m_N^2 \right)^2} \,  \left(1- \frac{m_\chi^2}{m_\phi^2}\right)^{3/2}\hspace{-0.5cm}, \\
\label{eq:sigvNN}
&\langle\sigma v\rangle_{\phi\phi\to N N} =\frac{\lambda^4 \left(m_\chi + m_N \right)^2}{2 \pi \left( m_\phi^2 + m_\chi^2 - m_N^2 \right)^2} \,  \left(1- \frac{m_N^2}{m_\phi^2}\right)^{3/2}\hspace{-0.5cm}, \\
\label{eq:PPhh}
&\langle\sigma v\rangle_{\phi\phi\to hh} = \left(1- \frac{m_h^2}{m_\phi^2}\right)^{1/2} \Bigg[ \frac{1}{64 \pi m_\phi^2}\times \nn \\ &\hspace{2.3cm}\left( 2\, \kappa + \frac{6 \,\kappa \,m_h^2 (4 m_\phi^2 - m_h^2)}{(4 m_\phi^2 - m_h^2)^2 + m_h^2 \Gamma_h^2}\right)^2 \nn \\
&\hspace{2.cm}+ \frac{\kappa^4 v^4 }{2 \pi m_\phi^2 \left( 2 m_\phi^2 -m_h^2 \right)^2} \Bigg], \\
\label{eq:PPSM}
&\langle\sigma v\rangle_{\phi\phi\to h\to {\rm SM}} = \frac{\sqrt{2}\kappa^2 v^2 G_F}{\pi\left((4 m_\phi^2 - m_h^2)^2 + m_h^2 \Gamma_h^2\right)} \times \nn \\ &\hspace{2.3cm}\Bigg[ 3 m_t^2\left(1- \frac{m_t^2}{m_\phi^2}\right)^{\!\!3/2}\!\!\!\! + 2 m_\phi^2\left(1- \frac{m_W^2}{m_\phi^2}\right)^{\!\!1/2}\!\!\!\!  \nn \\
&\hspace{2.2cm}+  m_\phi^2\left(1- \frac{m_Z^2}{m_\phi^2}\right)^{\!\!1/2}\Bigg].
\end{align}


\begin{thebibliography}{99}
	
	\bibitem{rhn}
	P. Minkowski, Phys. Lett. B67, 421 (1977);  
	T. Yanagida, in the Workshop 
	on Grand Unified Theory and Baryon Number of the Universe, KEK, Japan, 1979; 
	M. Gell-Mann, P. Ramond and R. Slansky in Sanibel Symposium, February 1979, CALT-68-709
	[retroprint arXiv:hep-ph/9809459], and in Supergravity, eds. D. Freedman et al. (North
	Holland, Amsterdam, 1979); 
	S. L. Glashow in Quarks and Leptons, Cargese, eds. M. Levy et. al. (Plenum, 1980, New York), p. 707; 
	R. N. Mohapatra and G. Senjanovic, Phys. Rev. Lett.
	44, 912 (1980).
	
	%\cite{Bandyopadhyay:2011qm}
	\bibitem{Bandyopadhyay:2011qm}
	P.~Bandyopadhyay, E.~J.~Chun and J.~C.~Park,
	%``Right-handed sneutrino dark matter in $\mathbf{U(1)'}$ seesaw models and its signatures at the LHC,''
	JHEP {\bf 1106} (2011) 129
	%doi:10.1007/JHEP06(2011)129
	[arXiv:1105.1652 [hep-ph]].
	%%CITATION = doi:10.1007/JHEP06(2011)129;%%
	%28 citations counted in INSPIRE as of 27 Feb 2018
	
	%\cite{Bandyopadhyay:2017bgh}
	\bibitem{Bandyopadhyay:2017bgh}
	P.~Bandyopadhyay, E.~J.~Chun and R.~Mandal,
	%``Implications of right-handed neutrinos in $B-L$ extended standard model with scalar dark matter,''
	Phys.\ Rev.\ D {\bf 97} (2018) no.1,  015001
	% doi:10.1103/PhysRevD.97.015001
	[arXiv:1707.00874 [hep-ph]].
	%%CITATION = doi:10.1103/PhysRevD.97.015001;%%
	%4 citations counted in INSPIRE as of 27 Feb 2018
	
	%\cite{Dror:2016rxc}\cite{Okawa:2016wrr} \cite{Kopp:2016yji}
	\bibitem{Dror:2016rxc}
	J.~A.~Dror, E.~Kuflik and W.~H.~Ng,
	%``Codecaying Dark Matter,''
	Phys.\ Rev.\ Lett.\  {\bf 117} (2016) no.21,  211801
	% doi:10.1103/PhysRevLett.117.211801
	[arXiv:1607.03110 [hep-ph]].
	%%CITATION = doi:10.1103/PhysRevLett.117.211801;%%
	%13 citations counted in INSPIRE as of 27 Feb 2018
	
	%\cite{Okawa:2016wrr}
	\bibitem{Okawa:2016wrr}
	S.~Okawa, M.~Tanabashi and M.~Yamanaka,
	%``Relic Abundance in a Secluded Dark Matter Scenario with a Massive Mediator,''
	Phys.\ Rev.\ D {\bf 95} (2017) no.2,  023006
	%doi:10.1103/PhysRevD.95.023006
	[arXiv:1607.08520 [hep-ph]].
	%%CITATION = doi:10.1103/PhysRevD.95.023006;%%
	%6 citations counted in INSPIRE as of 27 Feb 2018
	
	%\cite{Kopp:2016yji}
	\bibitem{Kopp:2016yji}
	J.~Kopp, J.~Liu, T.~R.~Slatyer, X.~P.~Wang and W.~Xue,
	%``Impeded Dark Matter,''
	JHEP {\bf 1612} (2016) 033
	%doi:10.1007/JHEP12(2016)033
	[arXiv:1609.02147 [hep-ph]].
	%%CITATION = doi:10.1007/JHEP12(2016)033;%%
	%12 citations counted in INSPIRE as of 27 Feb 2018
	
	\bibitem{posp07}
	%\cite{Pospelov:2007mp}
	%\bibitem{Pospelov:2007mp}
	M.~Pospelov, A.~Ritz and M.~B.~Voloshin,
	%``Secluded WIMP Dark Matter,''
	Phys.\ Lett.\ B {\bf 662} (2008) 53
	%  doi:10.1016/j.physletb.2008.02.052
	[arXiv:0711.4866 [hep-ph]].
	%%CITATION = doi:10.1016/j.physletb.2008.02.052;%%
	%507 citations counted in INSPIRE as of 01 Mar 2018
	
	
	
	\bibitem{falk09}
	%\cite{Falkowski:2009yz}
	%\bibitem{Falkowski:2009yz}
	A.~Falkowski, J.~Juknevich and J.~Shelton,
	%``Dark Matter Through the Neutrino Portal,''
	arXiv:0908.1790 [hep-ph].
	%%CITATION = ARXIV:0908.1790;%%
	%42 citations counted in INSPIRE as of 01 Mar 2018
	
	\bibitem{gonz16}
	%\cite{Gonzalez-Macias:2016vxy}
	%\bibitem{Gonzalez-Macias:2016vxy}
	V.~González-Macías, J.~I.~Illana and J.~Wudka,
	%``A realistic model for Dark Matter interactions in the neutrino portal paradigm,''
	JHEP {\bf 1605} (2016) 171
	% doi:10.1007/JHEP05(2016)171
	[arXiv:1601.05051 [hep-ph]].
	%%CITATION = doi:10.1007/JHEP05(2016)171;%%
	%12 citations counted in INSPIRE as of 01 Mar 2018
	
	\bibitem{esc16}
	%\cite{Escudero:2016ksa}
	%\bibitem{Escudero:2016ksa}
	M.~Escudero, N.~Rius and V.~Sanz,
	%``Sterile Neutrino portal to Dark Matter II: Exact Dark symmetry,''
	Eur.\ Phys.\ J.\ C {\bf 77} (2017) no.6,  397
	%  doi:10.1140/epjc/s10052-017-4963-x
	[arXiv:1607.02373 [hep-ph]].
	%%CITATION = doi:10.1140/epjc/s10052-017-4963-x;%%
	%14 citations counted in INSPIRE as of 01 Mar 2018
	
	\bibitem{tang16}
	%\cite{Tang:2016sib}
	%\bibitem{Tang:2016sib}
	Y.~L.~Tang and S.~h.~Zhu,
	%``Dark Matter Relic Abundance and Light Sterile Neutrinos,''
	JHEP {\bf 1701} (2017) 025
	%  doi:10.1007/JHEP01(2017)025
	[arXiv:1609.07841 [hep-ph]].
	%%CITATION = doi:10.1007/JHEP01(2017)025;%%
	%4 citations counted in INSPIRE as of 01 Mar 2018
	
	\bibitem{camp17}
	%\cite{Campos:2017odj}
	%\bibitem{Campos:2017odj}
	M.~D.~Campos, F.~S.~Queiroz, C.~E.~Yaguna and C.~Weniger,
	%``Search for right-handed neutrinos from dark matter annihilation with gamma-rays,''
	JCAP {\bf 1707} (2017) no.07,  016
	% doi:10.1088/1475-7516/2017/07/016
	[arXiv:1702.06145 [hep-ph]].
	%%CITATION = doi:10.1088/1475-7516/2017/07/016;%%
	%13 citations counted in INSPIRE as of 01 Mar 2018
	
	\bibitem{batell17}
	%\cite{Batell:2017rol}
	%\bibitem{Batell:2017rol}
	B.~Batell, T.~Han and B.~Shams Es Haghi,
	%``Indirect Detection of Neutrino Portal Dark Matter,''
	arXiv:1704.08708 [hep-ph].
	%%CITATION = ARXIV:1704.08708;%%
	%9 citations counted in INSPIRE as of 01 Mar 2018

	%\cite{Chianese:2018dsz}
	\bibitem{Chianese:2018dsz} 
	M.~Chianese and S.~F.~King,
	%``The Dark Side of the Littlest Seesaw: freeze-in, the two right-handed neutrino portal and leptogenesis-friendly fimpzillas,''
	arXiv:1806.10606 [hep-ph].
	%%CITATION = ARXIV:1806.10606;%%
	%1 citations counted in INSPIRE as of 23 Jul 2018
	
	%\cite{Athron:2018ipf}
	\bibitem{Athron:2018ipf}
	P.~Athron, J.~M.~Cornell, F.~Kahlhoefer, J.~McKay, P.~Scott and S.~Wild,
	%``Impact of vacuum stability, perturbativity and XENON1T on global fits of $\mathbb{Z}_2$ and $\mathbb{Z}_3$ scalar singlet dark matter,''
	arXiv:1806.11281 [hep-ph].
	%%CITATION = ARXIV:1806.11281;%%
	
	
	
	%\cite{Folgado:2018qlv}
	\bibitem{Folgado:2018qlv} 
	M.~G.~Folgado, G.~A.~Gómez-Vargas, N.~Rius and R.~Ruiz De Austri,
	%``Probing the sterile neutrino portal to Dark Matter with $\gamma$ rays,''
	arXiv:1803.08934 [hep-ph].
	%%CITATION = ARXIV:1803.08934;%%
	%2 citations counted in INSPIRE as of 03 Jul 2018
	
	\bibitem{Acharya:2017ttl} 
	B.~S.~Acharya {\it et al.} [Cherenkov Telescope Array Consortium],
	%``Science with the Cherenkov Telescope Array,''
	arXiv:1709.07997 [astro-ph.IM].
	%%CITATION = ARXIV:1709.07997;%%
	%57 citations counted in INSPIRE as of 05 Nov 2018
	
	
	%\cite{Ade:2015xua}
	\bibitem{Ade:2015xua} 
	P.~A.~R.~Ade {\it et al.} [Planck Collaboration],
	%``Planck 2015 results. XIII. Cosmological parameters,''
	Astron.\ Astrophys.\  {\bf 594}, A13 (2016)
	%  doi:10.1051/0004-6361/201525830
	[arXiv:1502.01589 [astro-ph.CO]].
	%%CITATION = doi:10.1051/0004-6361/201525830;%%
	%3325 citations counted in INSPIRE as of 16 May 2017
	
	%\cite{Aprile:2018dbl}
	\bibitem{Aprile:2018dbl}
	E.~Aprile {\it et al.} [XENON Collaboration],
	%``Dark Matter Search Results from a One Tonne$\times$Year Exposure of XENON1T,''
	arXiv:1805.12562 [astro-ph.CO].
	%%CITATION = ARXIV:1805.12562;%%
	%20 citations counted in INSPIRE as of 04 Jul 2018
	
	%\cite{Mount:2017qzi}
	\bibitem{Mount:2017qzi} 
	B.~J.~Mount {\it et al.},
	%``LUX-ZEPLIN (LZ) Technical Design Report,''
	arXiv:1703.09144 [physics.ins-det].
	%%CITATION = ARXIV:1703.09144;%%
	%11 citations counted in INSPIRE as of 26 Aug 2017
	
	%\cite{Aprile:2015uzo}
	\bibitem{Aprile:2015uzo} 
	E.~Aprile {\it et al.} [XENON Collaboration],
	%``Physics reach of the XENON1T dark matter experiment,''
	JCAP {\bf 1604}, no. 04, 027 (2016)
	%doi:10.1088/1475-7516/2016/04/027
	[arXiv:1512.07501 [physics.ins-det]].
	%%CITATION = doi:10.1088/1475-7516/2016/04/027;%%
	%223 citations counted in INSPIRE as of 26 Aug 2017
	
\end{thebibliography}
\end{document}